\documentstyle[11pt,newpasp,twoside,epsf]{article}
\def\edcomment#1{\iffalse\marginpar{\raggedright\sl#1\/}\else\relax\fi}
\reversemarginpar

\newcommand{\fig}[6]{
    \protect\centerline{
    \epsfxsize=#1\epsffile[#2 #3 #4 #5]{#6}
    }}

\begin{document}
\title{The Importance of Shocks in the Ionization of the Narrow Line Region
of Seyferts}
\author{H. R. Schmitt}
\affil{National Radio Astronomy Observatory, P.O. Box 0, Socorro, NM87801}
\author{A. L. Kinney}
\affil{NASA Headquarters, 300 E St., Washington, DC20546}
\author{J. B. Hutchings}
\affil{Dominion Astrophysical Observatory, Victoria, BC V8X 4M6, Canada}
\author{J. S. Ulvestad}
\affil{National Radio Astronomy Observatory, P.O. Box 0, Socorro, NM87801}
\author{R. R. J. Antonucci}
\affil{University of California, Santa Barbara, Santa Barbara, CA93106} 

\begin{abstract}

We discuss the viability of shocks as the principal source of ionization
for the Narrow Line Region of Seyfert galaxies. We present the preliminary
results of [OIII]$\lambda$5007\AA\ and radio 3.6~cm imaging surveys of
Seyferts, discuss the effects of shocks in the ionization of two galaxies,
and also calculate an upper limit to the H$\beta$ luminosity that can be due to
shocks in 36 galaxies with unresolved radio emission. We show that, for
favored values of the shock parameters, that shocks cannot contribute more than
$\sim$15\% of the ionizing photons in most galaxies. 

\end{abstract}

\section{Introduction}

The close connection between the narrow line and the radio emission in
Seyfert galaxies is a well known fact. The first papers in this
subject found a correlation between the [OIII] and radio luminosities,
and between the radio luminosity and the FWHM of the [OIII] lines
(DeBruyn \& Wilson 1978, Heckman et al. 1981, Whittle 1985).
Emission line and radio imaging of Seyfert galaxies shows that
the radio and line emission of these galaxies is aligned (Haniff et al. 1988).
Higher resolution imaging shows, for some of these galaxies, that the
radio emission  is surrounded by line emission (Pogge \& De Robertis 1995,
Capetti et al. 1996, Falcke et al. 1998), which suggests that the 
Narrow Line Region gas (NLR) is ionized by shocks due to the interaction
of the radio plasma with the gas (see Figure 1 for an example).
Another evidence of shocks in the NLR of Seyferts is the
detection of anomalous velocity fields, like double components separated by
hundreds of km/s (Whittle et al. 1988, Winge et al. 1997,
Hutchings et al. 1998), and ``extra'' line widths associated with radio
jets (Whittle 1992).

These results lead Dopita \& Sutherland (1995, 1996) to propose a series of
models where the NLR is ionized by fast shocks, with velocities in the range
150$\leq V_{shock} \leq 500$ km/s. According to their models, high energy
photons are created behind the shock and ionize the preshock gas upstream
(precursor). This model has several predictions, which were analyzed
by Morse et al. (1996) and Wilson (1997), and are summarized below.
The models predict we should find line emission surrounding
radio lobes, and a photon deficit in the NLR. However,
this can also be explained by the Unified Model (see Young et al. 2001
for a discussion on X-ray results). Shocks can explain
the [OIII] temperature problem in these galaxies, where the measured gas
temperature is much higher than the one predicted by simple photoionization
models, but this can
also be explained combining matter bounded and ionization bounded
clouds (Binette et al. 1996).

Here we present the preliminary results of a radio and [OIII]$\lambda$5007\AA\
imaging survey of a sample of Seyfert galaxies. We discuss the case of two
individual galaxies and calculate a limit to the H$\beta$ luminosity that
can be due to shocks in those galaxies without extended radio emission.

\begin{figure}
    \fig{12cm}{40}{230}{600}{500}{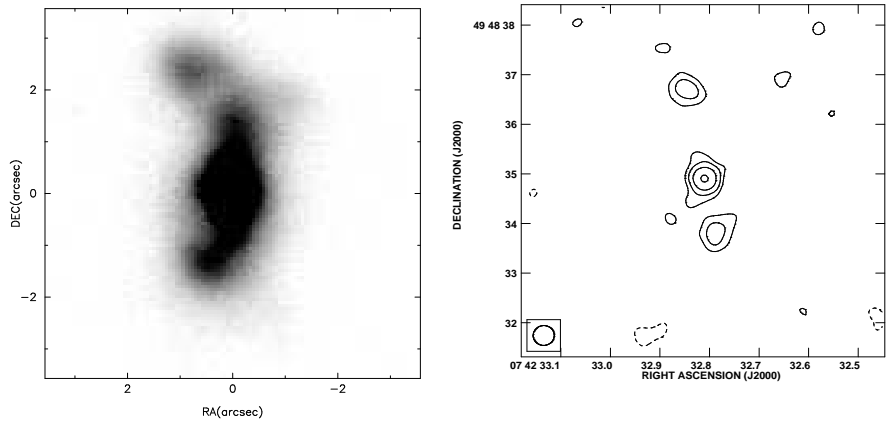}
\caption{Comparison of the CFHT [OIII]$\lambda$5007\AA\ image of Mrk\,79
(left) with the VLA A-conf. 3.6~cm image (right), on the same scale.}
\end{figure}

\section{Radio and [OIII] survey of Seyfert galaxies}

Our radio continuum survey was done using the VLA in A-configuration at
3.6~cm, which gives a resolution of 0.25$^{\prime\prime}$. The results of
this survey were published by Kinney et al. (2000) and Schmitt et al. (2001).
The [OIII] survey is underway. It started with the CFHT and is now being
done with HST.

One of the results from the radio continuum survey is that only 50\%
of the Seyferts show extended emission. This percentage decreases to 25\%
if we count only those galaxies where the radio emission is composed of multiple
components. Comparing the radio and [OIII] images we find some interesting
cases, like NGC5347 (Figure 2). This Figure shows that the [OIII] emission
is resolved, but the radio emission is unresolved, contrary to what would be
expected if shocks were important in the ionization of the NLR of this
galaxy. On the other hand we also have cases like Mrk\,79 (Figure 1), where
both the [OIII] and radio emission are extended and cospatial, suggesting
that shocks may be an important source of ionizing photon.
We discuss these galaxies below.

We will use in the following subsections the equations given by Dopita \&
Sutherland (1996) to predict the H$\beta$ luminosity produced by a shock
and the precursor, which are, respectively,
$$ L(H\beta) = 7.44\times10^{-6} A (V_s/100 {\rm ~km~s^{-1}})^{2.41} (n/{\rm ~cm^{-3}}) {\rm ~erg~s^{-1}}$$
$$ L(H\beta) = 9.85\times10^{-6} A (V_s/100 {\rm ~km~s^{-1}})^{2.28} (n/{\rm ~cm^{-3}}) {\rm ~erg~s^{-1}}$$
\noindent
where $A$ is the shock area, $V_s$ is the shock velocity and $n$
is the preshock density.

\subsection{NGC5347}

This is a Seyfert 2 galaxy with a radial velocity of 2335 km s$^{-1}$,
so 1$^{\prime\prime}$ corresponds to 150 parsecs at the galaxy
(we assume H$_0=75$ km s$^{-1}$ Mpc$^{-1}$).
As we can see in Figure 2, the
[OIII] image is extended, with a cone shaped structure in the circumnuclear
region. At $\approx3^{\prime\prime}$ NE from the nucleus we can see [OIII]
emission detached from the nuclear emission (NE structure), with a
total linear extent of 2$^{\prime\prime}$. The borders of this structure are
well aligned with the borders of the cone structure at the nucleus, suggesting
that the NLR is ionized by the nuclear source, and the collimation of the
radiation is due to a circumnuclear torus. Figure 2 also shows the
archival VLA 6~cm image of this galaxy, which is unresolved. The radio emission
is unresolved at 3.6~cm, 6~cm and 20~cm, indicating that we did not resolve
out diffuse emission around the nucleus. In fact, the comparison between the
20~cm flux from the nuclear source (3.4~mJy) with that from the entire galaxy
(5.6~mJy), obtained from the NVSS, shows that most of the radio emission
in this galaxy originates in the nuclear region.

Here we calculate how much of the H$\beta$ emission from the NE structure can
be due to shocks. If we assume that the H$\beta$ flux
is produced by a shock seen edge-on,
with a circular cross section of radius 1$^{\prime\prime}$, we get that the
shock area is 0.071 kpc$^2$ (6.75$\times10^{41}$ cm$^2$). Gonz\'alez Delgado \&
P\'erez (1996) obtained spectra of this galaxy and showed that the
NE structure has L(H$\beta)=2\times10^{39}$ erg~s$^{-1}$, [OIII]/H$\beta=7.6$,
$n_e=350$~cm$^{-3}$, and the FWHM([OIII])$=80$ km~s$^{-1}$.
According to Ferruit et al. (1999), the pre-shock density for a shock with
this velocity would be of the order of $n=15$~cm$^{-3}$. Using the equations
given above, for a shock of velocity
$V_s=100$~km~s$^{-1}$ and preshock density $n=15$~cm$^{-3}$, we get
L(H$\beta)=7.5\times10^{37}$ erg~s$^{-1}$ and L(H$\beta)=1.0\times10^{38}$
erg~s$^{-1}$ for the shock and precursor, respectively. This corresponds to
$\approx$10\% of the observed H$\beta$ luminosity. Another problem with the
shock ionization of the NE structure is that the [OIII]/H$\beta$ ratio
can only be reproduced with a shock of 400 km~s$^{-1}$,
which is inconsistent with the observed FWHM of the [OIII] line.

We now use this same approach for the nuclear region of this galaxy.
According to Gonz\'alez Delgado \& P\'erez (1996), the nuclear spectrum
has L(H$\beta)=1.0\times10^{40}$ erg~s$^{-1}$ and
$n_e=400$~cm$^{-3}$. They also found that the [OIII] line can be decomposed
into a narrow and a broad component, with the broad component having
FWHM([OIII])=500~km~s$^{-1}$. Again, if we assume that we are seeing
the shock edge-on, with a circular cross section of radius
0.5$^{\prime\prime}$, we get that the shock area is 0.018~kpc$^2$
(1.7$\times10^{41}$ cm$^2$). We get that for a shock velocity
$V_s=500$~km~s$^{-1}$ and a preshock density $n=10$~$cm^{-3}$,
L(H$\beta)=6.3\times10^{38}$ erg~s$^{-1}$ for the shock and
$6.6\times10^{38}$ erg~s$^{-1}$ for the precusor, which corresponds to
$\approx10$\% of the observed H$\beta$ luminosity. These results
indicate that shocks are not extremely important for the ionization of the
NLR of this galaxy.

\begin{figure}
    \fig{12cm}{40}{230}{600}{500}{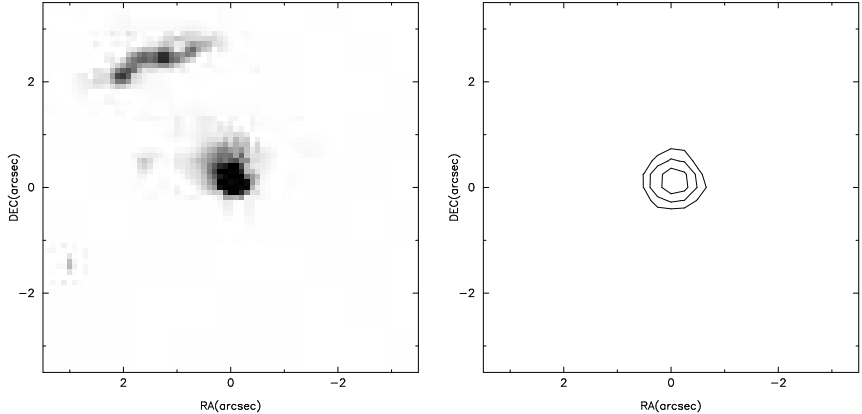}
\caption{Comparison between the HST [OIII]$\lambda$5007\AA\ images of NGC\,5347
(left) with the archival VLA A-config. 6~cm image (right).}
\end{figure}

\subsection{Mrk\,79}

The [OIII] and radio images of Mrk\,79 are shown in Figure 1. This Figure
shows that the [OIII] emission of this Seyfert 1 galaxy is extended by
$\approx5^{\prime\prime}$ along the N-S direction. The radio emission
lies along the same direction, but is extended by only
$\approx3^{\prime\prime}$. The radio structures to the N and S of the nucleus
are surrounded by the [OIII] emission, suggesting that shocks may play an
important role in the ionization of the gas. Using the same technique used
for NGC5347, we calculate what would be the contribution of shocks to the
H$\beta$ luminosity of the N and S regions of the NLR. We assume the shocks
have a circular cross section of radius 0.5$^{\prime\prime}$, which
corresponds to an area of 0.145~kpc$^2$ ($1.4\times10^{42}$ cm$^2$),
considering the galaxy has a radial velocity of 6643~km~s$^{-1}$. From Whittle
et al. (1988) we have that the N and S lobes have FWHM([OIII])=400~km~s$^{-1}$
and 350~km~s$^{-1}$, respectively, which we assume to be the shock velocities.
Assuming a preshock density $n=$10~cm$^{-3}$, we can calculate that the
sum of the shock and precursor H$\beta$ luminosities are
$6.1\times10^{39}$~erg~s$^{-1}$ for the N component, and
$4.5\times10^{39}$~erg~s$^{-1}$ for the S component. Using our CFHT
H$\alpha$ image and assuming H$\alpha$/H$\beta=3.1$, we get
the observed L(H$\beta$)$=2.8\times10^{40}$~erg~s$^{-1}$ and
$1.9\times10^{40}$~erg~s$^{-1}$ for the N and S components, respectively.
This means that $\approx$25\% of the photons ionizing the N and S
lobes of this galaxy can be due to shocks.

\subsection{Seyferts with unresolved radio emission}

Another test of the importance of shocks to the ionization of the NLR
of Seyferts was done using those galaxies for which we do not observe
extended radio emission. We use the same method used for NGC5347
and Mrk\,79. Since these galaxies are unresolved in radio, we make the
conservative assumption that the shock has a circular cross section with 
radius 0.25$^{\prime\prime}$, two times larger than the resolution
of the observations. We do not have information about the FWHM([OIII])
of these galaxies, so we assume that all the galaxies have shocks of
500 km~s$^{-1}$, at the high end of the parameters modeled by Dopita
\& Sutherland (1995, 1996), and use a preshock density of $n=10$~cm$^{-3}$.
Given these assumptions, we should consider these results as upper limits to
the contribution of shocks to the ionization of these NLR's.

\begin{table}
\caption{Observed and predicted H$\beta$ luminosities$^{1}$}
\begin{tabular}{lrrr|lrrr}
\tableline
Name& L(H$\beta$)& L(H$\beta$)&Ref.&
Name& L(H$\beta$)& L(H$\beta$)&Ref.\cr
    & Obs        & Calc       &    &
    & Obs        & Calc       &    \cr  
\tableline
Mrk\,1&  16.55& 1.34&1       & I\,01475-0740&   5.99& 1.65&1\cr
Mrk\,1040&   7.24& 1.42&2    & UGC\,2024&   6.72& 2.64&1\cr
MCG\,-2-8-39&  20.13& 4.61&1 & I\,03125-0254&  10.93& 3.03&1\cr
Mrk\,607&   2.20& 0.43&1     & I\,04385-0828&  1.27& 1.20&1\cr
MCG\,-5-13-17&  3.37& 0.84&3 & UGC\,3478&  1.87& 0.86&1\cr
UGC\,4155& 26.14& 3.42&1     & Mrk\,1239& 39.53& 2.09&3\cr
NGC\,3783&  4.99& 0.38&4     & NGC\,4593& 10.23& 0.43&5\cr
NGC\,4704&  8.46& 3.87&1     & MCG\,-2-33-34& 19.67& 1.13&3\cr
I\,13059-2407&  1.49& 1.02&1 & MCG\,-6-30-15&  1.10& 0.32&6\cr
NGC\,5347& 12.00&1.47&11     & I\,14434+2714& 31.44& 4.54&1\cr
UGC\,9826&  8.49& 4.48&1     & I\,15480-0344& 10.55& 4.83&1\cr
I\,16288+3929& 26.41& 4.83&1 & I\,16382-0613& 11.64& 4.05&1\cr
UGC\,10889&  7.56& 4.15&1    & MCG\,+3-45-3& 30.58& 3.11&1\cr
UGC\,11630&  1.22& 0.78&1    & NGC\,7213&  3.69& 0.19&7\cr
Mrk\,915& 35.63& 3.06&2      & UGC\,12138& 16.22& 3.18&1\cr
UGC\,12348& 26.96& 3.37&1    & Mrk\,590& 22.66& 3.66&8\cr
Mrk\,1058 &  1.74& 1.54&2    & Mrk\,705& 19.16& 4.39&9\cr
UGC\,6100& 39.39& 4.51&9     & UGC\,10683\,B&  6.54& 4.99&10\cr
\tableline
\end{tabular}
\footnote{}{Luminosities are in units of 10$^{39}$~erg~s$^{-1}$.
The H$\beta$ luminosities calculated based on shock models include
both the shock and the precursor. References: (1) de Grijp et al. (1992);
(2) Dahari \& De Robertis (1988); (3) Rodr\'{\i}guez-Ardila et al. (2000);
(4) Winge et al. (1992); (5) Clavel et al. (1983); (6) Reynolds et al.
(1997); (7) Filippenko et al. (1988); (8) Stirpe (1990); (9)
Cruz-Gonzalez et al. (1994); (10) Wilson et al. (1981); (11) Gonz\'alez Delgado
\& P\'erez (1996).}
\end{table}

We show in Table 1 the observed and calculated (shock$+$precusor)
H$\beta$ luminosities of the galaxies with unresolved radio emission,
as well as the references from which we obtained the observed values.
On average only 15\% of the observed H$\beta$ luminosity can be due to shocks,
and this number is likely to be smaller, given the favorable shock parameters
we assumed.

\section{Summary}

We presented the preliminary results of a survey of radio and [OIII] images
of Seyfert galaxies. About $\sim$50\% of our galaxies have
unresolved radio emission (smaller than 0.25$^{\prime\prime}$).
Based on conservative assumptions about the area and velocity of
a shock in these galaxies, and on the preshock gas density, we calculated
the shock and precursor H$\beta$ luminosities for each one of them and compared
these values with the observed ones. This showed that, on average,
only 15\% of the observed H$\beta$ emission can be due to shocks, which is
confirmed by the detailed study of two individual galaxies with resolved
[OIII] and radio emission (Mrk\,79 and
NGC\,5347). In summary, in most of the cases shocks are not a viable source
of ionizing photons for the NLR of Seyfert galaxies.

\acknowledgements

Support for this work was provided by NASA grants AR-8383.01-97A and
GO-08598.07-A.
The National Radio Astronomy Observatory is a facility
of the National Science Foundation operated under cooperative agreement
by Associated Universities, Inc.

\end{document}